\documentclass[11pt]{article}

\usepackage{graphics}   
\usepackage{amssymb}
\usepackage{amsmath}
\usepackage{amsthm}
\usepackage{float}

\newtheorem{theorem}{Theorem}
\newtheorem{corollary}{Corollary}

\theoremstyle{definition}

\newtheorem{remark}{Remark}

\def\noi{\noindent}

\def\parsec{\par\noindent}

\def\med{\medskip\parsec}
\def\E{\mathbb{E}}

\def\A{\mathcal{A}}

\def\noi{\noindent}

\def\parsec{\par\noindent}

\def\med{\medskip\parsec}

\def\R{\bar{R}}

\def\th{\theta}

\def\E{{\bf E}}

\def\a{\alpha}

\def\A{{\cal A}}

\def\S{{\cal S}}

\def\bx{{\bf x}}
\def\ba{{\bf a}}
\def\bb{{\bf b}}

\def\bar{\overline}
\def\tilde{\widetilde}

\def\Dir{{\rm Dir}}

\def\veca{\mbox{\boldmath $\alpha$}}
\def\vecth{\mbox{\boldmath $\theta$}}
\def\M{{\cal M}}

\def\1/2{{\bf 1/2}}

\begin{document}
\begin{center}
{\Large {\bf Sequential Universal Modeling for Non-Binary Sequences
with Constrained Distributions}}
\parsec
\end{center}
\med
\par
\noi

\begin{center}
\begin{tabular}{lll}
Michael Drmota&Gil I. Shamir&Wojciech Szpankowski\footnotemark \\
Institut f\"ur Geometrie&Google,Inc&Dept. of Computer Science\\
TU Wien&&Purdue University\\
A-1040 Wien&Pittsburgh, PA&W. Lafayette, IN 47907\\
Austria&U.S.A.&U.S.A.\\
{\tt drmota@tuwien.ac.at}&{\tt gshamir@ieee.org}&{\tt szpan@purdue.edu}
\end{tabular}
\end{center}

\footnotetext{
M. Drmota was supported in part by the
the grant FWF Grant SFB F50-02.
W. Szpankowski was supported in part by
NSF Center on Science of Information
Grants CCF-0939370 and NSF Grants CCF-1524312, CCF-2006440, CCF-2007238,
and Google Research Award.}

\medskip 

\begin{abstract}
Sequential probability assignment and universal compression go hand in hand.
We propose sequential probability assignment for
non-binary (and large alphabet) sequences with 
distributions whose parameters are known to be bounded
within a limited interval. Sequential probability assignment algorithms
are essential in many applications that require fast and accurate
estimation of the maximizing sequence probability. These applications include
learning, regression, channel estimation and decoding, prediction,
and universal compression. On the other hand, constrained
distributions introduce interesting theoretical twists that
must be  overcome in order to present efficient sequential algorithms.
Here, we focus on universal compression for memoryless sources, and present a precise
analysis for the maximal minimax and the (asymptotic)
average minimax redundancy for constrained distributions. We show that our
sequential algorithm based on modified Krichevsky-Trofimov (KT) estimator
is asymptotically optimal  up to $O(1)$ for both maximal and average redundancies.
In addition, we provide precise asymptotics of the minimax 
redundancy for monotone distributions
which is a special case of the constrained distribution.
This paper follows and addresses the challenge presented in \cite{stw08} that
suggested `results for the binary case lay
the foundation to studying larger alphabets''.
\end{abstract}
\med

\section{Introduction}

Universal coding and universal modeling (probability assignments) 
are two driving forces of
information theory, model selection, and statistical inference.
In universal coding one is to construct a code for data sequences
generated by an unknown source from a known family such that, as the length of the sequence
increases, the average code length approaches the entropy of whatever
processes in the family has generated the data. In seminal works of
Davisson \cite{davisson73}, Rissanen \cite{rissanen96},
 Krichevsky and Trofimov \cite{kt83}, and 
Shtarkov \cite{shtarkov87}
it was shown how to construct such codes for finite alphabet sources.
Universal codes are often characterized by the average {\it minimax}
redundancy which is the excess over the entropy of the {\it best} code
from a class of decodable codes for the worst process in the family.

As pointed out by Rissanen \cite{rissanen96},
over years universal coding evolved into {\it universal modeling}
where the purpose is no longer restricted to just coding but rather to
learn optimal models \cite{rissanen96}. The central question
of interest in universal modeling seems to be in universal codes achievable
for {\it individual} sequences. The burning question is how to measure its performance.
The {\it worst case} minimax redundancy  became handy since it
measures the worst case excess of the best code maximized over
the processes in the family.
Unfortunately, low-complexity universal codes that are optimal for 
the worst case minimax are not easily implementable (it would require to
implement the maximum likelihood distribution). Therefore, we design a sequential
algorithm based on the KT-estimator that is asymptotically optimal on average
(i.e., for the average minimax redundancy), 
and show that both redundancies differ by a small constant.  

In this paper we focus on universal compression and probability 
assignment/learning for
a class of memoryless sources with {\it constrained distributions}. 
Let us start with
some definitions and notation. We define a 
code $C_n:{\cal A}^n \to \{0,1\}^*$ as a mapping from the
set $\A^n$ of all sequences $x^n = (x_1,\ldots,x_n)$
of length $n$ over the finite alphabet $\A=\{1, \ldots, m\}$ of size $m$ to the
set $\{0,1\}^*$ of all binary sequences.
Given a probabilistic source model, we let $P(x^n)$ be the
probability of the message $x^n$;
given a code $C_n$, we let $L(C_n, x^n)$ be the code length for $x^n$.
However, in practice the probability distribution
(i.e., source) $P$ is unknown, and one looks for
{\it universal codes} for which
the redundancy is $o(n)$ for all $P\in\S$ where $\S$ is a class of source
models (distributions). It is convenient to 
ignore the integer nature of the code length and replace it  by 
its best distributional guess, say $Q(x^n)$. In other words, we just
write $L(C_n,x^n)=-\log Q(x^n)$ and use it throughout the paper.
The question is how well $Q$ approximates $P$ within the class $\S$. 
Minimax redundancy enters. Usually, we consider two types of minimax redundancy,
namely {\it average} and {\it maximal} or {\it worst case} defined,
respectively, as
\begin{align}
\R_n(\S)&=\min_{Q }\sup_{P\in \S} {\bf \E} [\log P(X^n)/Q(X^n) ] , \label{e1}\\
R_n^*(\S)&=\min_{Q}\sup_{P\in \S} \mbox{\boldmath $\max$}_{x^n}[\log P(x^n)/Q(x^n) ].
\label{e2}
\end{align}

In this paper we analyze precisely both redundancies 
for {\it memoryless sources} over $m$-ary alphabet $\A=\{1, \ldots, m\}$
with {\it restricted symbol probability} $\th_i$, that is,
we assume that $\vecth \in \S$, where $\S$ is a proper subset of
\[
\Theta = \{ \vecth : \th_i \ge 0 \ (1\le i\le m),\  \theta_1 + \cdots + \theta_m = 1\}.
\]
We will assume that $\S$ is a convex polytope. 
As a special case we have the {\it interval} restriction
$0\le a_i\leq \theta_i \leq b_i\le 1$ for $i=1, \ldots m-1$,
where $\sum_{i=1}^{m-1} b_i \le 1$ (this ensures that $\theta_m = 1 - \sum_{i=1}^{m-1}$ 
is always well defined).
Also, a class of monotone distributions \cite{shamir13} defined as
$$
\M=\{ \vecth : 0\le \th_1\le \th_2 \le \cdots \le \th_m ,\  \theta_1 + 
\cdots + \theta_m = 1\}.
$$
is a special case of the constrained distributions.

Here, we present a sequential algorithm that estimates asymptotically
the optimal probability  $P(x^n)$ for all $x^n$. 
It turns out that restricting the set of parameters 
is important from a practical point of view
and at the same time introduces new interesting theoretical
twists that we explore in this paper.
We first prove in Theorem~\ref{th-r*} that (for fixed $m$ that can still be large)
\begin{align*}
\R_n(\S) = R_n^*(\S) + O(1) 
= \frac{m-1}{2} \log (n) + O(1)
\end{align*}
where the constant implied by the $O$-term depends on $m$ and on the set $\S$.
Second we provide in Theorem~\ref{th-rbar} precise asymptotics 
for $\R_n(\Theta)$ and $R_n^*(\Theta)$ 
if $m= o(n)$. While the leading terms of these redundancies are known 
\cite{alon04,shamir06,shamir06b,sw12}, 
we derive here {\it precise} asymptotics up to 
$O(m^{3/2}/\sqrt{n})$ term in a uniform manner
that can be used to extend our analysis to the constrained case in this regime. 
This allows us in Theorem~\ref{cor1} to provide the best asymptotic expansions
for the minimax redundancy of monotone distributions.
Finally, we present in Corollary~\ref{th-algo} a sequential 
add-1/2 KT-like estimator to 
compute $P(x_{n+1}|x^n)$ for the constrained distributions that is
asymptotically optimal up to a constant
for both the maximal and average redundancy. 
This final result has been wanting since \cite{stw08} which suggested that
``results for the binary case lay
the foundation to studying larger alphabets''.

This paper is organized as follows. In the next section we present our main results,
including the sequential algorithm that directly generalizes the
add-1/2-KT estimator. The proofs are discussed in the last section and 
an appendix.

\section{Main Results}

In this section we present our main results including asymptotically optimal
probability estimation for the class $\S\subset \Theta$ of memoryless sources with
constrained distributions.

We start with the {\it worst case redundancy} defined in (\ref{e2}). We recall that
the distribution  $P(x^n)$ is of the form
$$
P(x^n)=\prod_{i=1}^m \th_i^{k_i}, \ \ \ \ \ \  \th_i \ge 0,\ \sum_{i=1}^m \th_i = 1,
$$
where $k_i$ is the number of symbol $i\in \A$ in the sequence $x^n$. 
The probabilities $\th_i$ are unknown to us except that we restrict them to the
subset $\S\subseteq \Theta$. Following Shtarkov \cite{shtarkov87} and \cite{ds04}
we can re-write the worst case redundancy for $\S$,
by noting that $\max$ and $\sup$ commute, as
\begin{align*}
R_n^*(\S) &= \min_{Q}\sup_{P\in \S}\max_{x^n}
(-\log Q(x^n)+\log P(x^n)) \\
&= \min_{Q} \max_{x^n} [-\log Q(x^n)+
\sup_{P\in \S} \log P(x^n) ] \\
&= \min_{Q} \max_{x_n} [\log Q^{-1}(x^n) 
 +\log P^*(x^n) + \log \sum_{z^n} \sup_{P\in \S} P(z^n) ] \\ 
& = \log \sum_{x^n} \sup_{P\in \S} P(x^n) 
\end{align*}
where $P^*(x^n)$ is
\begin{equation}
\label{eq-p}
P^*(x^n) := \frac {\sup\limits_{P\in \S} P(x^n)}{\sum_{z_1^n} \sup_{P\in \S} P(z_1^n)}
\end{equation}
is the {\it maximum-likelihood distribution} and we set $Q(x^n) = P^*(x^n)$
for attaining the minimum. In this context the distribution $P^*$ is also called
Shtarkov distribution and the sum 
\[
D_n = \sum_{x^n} \sup_{P\in \S} P(x^n) 
\]
is called Shtarkov sum. Note that $R_n^*(\S) = \log D_n$.

If we define the worst case redundancy with the help of code lengths $L(C_n,x^n)$ 
instead of $-\log Q(x^n)$ -- that we denote by $\tilde R_n^*$ -- 
then we would get a similar expression of the
form $\tilde R_n^*(\S) = \log D_n + \tilde R_n^*(P^*)$.
Using Shannon's code Shtarkov immediately noticed that $0<\tilde R_n^*(P^*)<1$, and 
in \cite{ds04} it was actually proved that asymptotically  
for the unconstrained case
$$
\tilde R_n^*(P^*)={\log \left(\frac{1}{m-1} \log  m\right) \over \log m} +o(1).
$$
From now on we shall ignore this {\it correction term} 
and analyze $R_n^*(\S) = \log D_n$.

To estimate the sum 
$D_n=\sum_{x^n} \sup_P P(x^n)$ 
we need first to find $\sup \prod_{i=1}^m \th_i^{k_i}$
when $\vecth\in \S$. For the unrestricted case ($\S=\Theta$) we know
that the optimal $\th_i=k_i/n$. The situation is more complicated in the
constrained case. For example, if we assume an interval 
restriction $a_i \le \th_i \le b_i$,
$i=1,\ldots, m-1$ with $\th_m=1-\th_1 - \cdots - \th_{m-1}$, 
then for  $k_i< n a_i$ or $k_i> n b_i$ the optimal $\th_i$ may 
be $a_i$ or $b_i$, respectively. 
Fortunately, we are able to prove that the main contribution to $D_n$ comes from
those ${\bf k} = (k_1,\ldots, k_m)$ for which ${\bf k}/n \in \S$
(see (\ref{eq-error}) and Appendix). 
So we are led to analyze the following sum
$$
D_n^{(\S)} =\sum_{ {\bf k} \in n \S} {n \choose k_1, \cdots k_m} \prod_{i=1}^m 
\left(\frac{k_i}{n}\right)^{k_i}
$$
which is of order $n^{\frac{m-1}{2}}$. 
The contribution of the remaining terms is typically of order $O(n^{\frac{m-2}{2}})$. 

It is our goal to present a sequential low-complexity algorithm for the probability
assignment, that is, an iterative procedure to compute $P(x_{n+1}| x^n)$.
Unfortunately, the maximum-likelihood distribution (\ref{eq-p}) 
is not well suited for it.
To find one, we switch to the {\it average} minimax redundancy (\ref{e1}) and we
re-cast in the Bayesian framework. 

Before we discuss the average minimax redundancy, we need to introduce
one more notation element. Let us define the Dirichlet density  as
$$
\Dir(\theta_1, \ldots, \theta_m; \alpha_1, \ldots,  \alpha_m)=
\frac{1}{B(\alpha_1, \ldots, \alpha_m)}  \prod_{i=1}^m \theta_i^{\alpha_i -1},
$$
where
$\sum_{i=1}^m \theta_i=1$ and
$$
B(\alpha_1, \ldots, \alpha_m)= B_m(\alpha_1, \ldots, \alpha_m) =  \frac{\Gamma(\alpha_1) \cdots \Gamma(\alpha_m)}{
\Gamma(\alpha_1+ \cdots + \alpha_m)}$$
is the beta function.
We shall write $\veca=(\a_1, \ldots \a_m)$ and $\vecth=(\theta_1, \ldots \theta_m)$ with
$\sum_{i=1}^m \theta_i=1$. Finally, we set for $\S \subset \Theta$
$$
\Dir(\S; \veca)=\frac{1}{B(\veca)} \int_{\S} \vecth^{\veca -1}\, d{\vecth}.
$$

\def\Tab{{\Theta_{\ba,\bb}}}
\def\Cab{{C_{\ba,\bb}}}
\def\bk{{\bf k}}

Let $\S \subseteq \Theta$. Then 
the {\it average minimax} problem is 
$$
\R_n(\S)=\inf_Q \sup_{\th \in \S} D_n(P^\th \| Q)
$$
where $D(P^\th \| Q)$ is the Kullback-Leibler divergence.
In the Bayesian framework, one assumes that the parameter $\th$  is generated by
the density $w(\th)$ and the mixture $M_n^w(x^n)$ is 
$$
M_n^w(x^n)=\int_\S P^\th(x^n) w(d\th).
$$
Observe now
\begin{align*}
\inf_Q \E_w[D_n(P^\th \| Q)] &=
\inf_Q\int_{\S} D_n(P^\th \| Q) dw(\th) \\
&= \int_{\S} D_n(P^\th \| M_n^w) dw(\th),
\end{align*}
where we use the fact that 
$$\min_Q \sum_i P_i \log 1/{Q_i}=\sum_i P_i \log 1/{P_i}.$$
As pointed out by Gallager \cite{gallager}, and Davisson \cite{davisson73} 
the minimax theorem of game theory
entitles us to conclude that
$$
\R_n(\S)=\inf_Q \sup_{\theta\in \S} D_n(P^\th \| Q) =
\sup_w \inf_Q \E_w[D_n(P^\th \| Q)]
$$
leading to
\begin{equation}
\label{eq-r2}
\R_n(\S)=\int_{\S} D(P^\th \| M_n^{w^*}) dw^*(\th)
\end{equation}
where $w^*(\th)$ is the maximizing prior distribution. 
Bernardo \cite{bernardo} proved that asymptotically
the maximizing density is proportional to the square root of the determinant of the
Fisher information $I(\vecth)$, the so-called Jeffrey prior. This leads to the density
\begin{equation}
\label{eq-w1}
\tilde w^*(\vecth)= \frac{1}{C(\S) \cdot B({\bf 1/2})} 
\frac{1}{\sqrt{\th_1 \cdots \th_m}},
\end{equation}
where $C(\S)$ defined as
\begin{equation}
\label{eq-cab}
C(\S)=\Dir(\S; {\bf 1/2})=
\frac{1}{B({\bf 1/2})} \int_{\S} \frac{d\vecth}{\sqrt{\th_1 \cdots \th_m}}
\end{equation}
is the probability that the Dirichlet distribution with $\alpha_i=1/2$ falls into
the subset $\S$.  
For example, Clarke and Barron \cite{clarkebarron} showed 
(under proper regularity conditions are satisfied, including the 
finiteness of the determinant of Fisher information and that $\S$ is a 
compact subset of the interior of $\Theta$) that
\begin{align*}
&\lim_{n\to\infty} \left( \int_{\S} D(P^\th \| M_n^{w^*}) dw^*(\th) - \frac{m-1}2\log \frac n{2\pi e} \right) \\
& = \lim_{n\to\infty} \left( \int_{\S} D(P^\th \| M_n^{\tilde w^*}) d\tilde w^*(\th) - \frac{m-1}2\log \frac n{2\pi e} \right) 
= \log \int_{\S} \sqrt{\det I(\vecth)}\, d\vecth.
\end{align*}
Barron and Xie \cite{xb00} extended this result to the 
unconstrained case $\S = \Theta$.
We note that $C(\Theta)=1$ for the unconstrained case.

This leads us to the following notation of the {\it asymptotic 
average minimax redundancy}
$$
\R_n^{\rm asymp}(\S)=\int_{\S} D(P^{\vecth} \| M_n^{\tilde w^*}) d \tilde w^*(\vecth).
$$
The mixture distribution $M_n^{\tilde w^*}(x^n)$ can be calculated as follows
\begin{align}
\nonumber
M_n^{\tilde w^*}(x^n)&= \frac{1}{C(\S) \cdot B({\bf 1/2})} \int_{\S} \prod_{i=1}^m 
\theta_i^{k_i-1/2} \\
\nonumber
&= \frac{1}{C(\S) \cdot B({\bf 1/2})} B(k_1+1/2, \cdots , k_m+1/2) \\ 
\nonumber
&\cdot \frac{1}{B(k_1+1/2, \cdots , k_m+1/2)} \int_{\S} \prod_{i=1}^m 
\theta_i^{k_i-1/2} \\
&= \frac{1}{C(\S) \cdot B({\bf 1/2})} B(k_1+1/2, \cdots , k_m+1/2) 
\cdot  \Dir(\S: \bk+{\bf 1/2}) 
\label{eq-m2}.
\end{align}
Observe again that for the unconstrained case $\Dir(\Theta; \bk+1/2)=1$.
In summary
\begin{align}
\label{eq-d1}
D_n(P^\th \| M_n^{\tilde w^*}) &=
\log \left(C(\S) B({\bf 1/2}) \right) +\\
&+ \sum_{\bk}{n \choose \bk}\prod_{i=1}^m \th_i^{k_i} 
\log \frac{\prod_{i=1}^m \th_i^{k_i}}
{B(\bk+{\bf 1/2}) \Dir(\S; \bk+{\bf 1/2})}
\end{align}

We are now ready to formulate our first main result that reads as follows.
We prove them in the next section and delay some technical derivations to Appendix.

\begin{theorem}
\label{th-r*}
Consider a memoryless constrained source $\S \subset \Theta$ with
fixed but arbitrarily large $m\ge 2$ where $\S$ is a convex polytope.
Then the worst case minimax redundancy for $\S$ is
\begin{align}
\nonumber
R_n^*(\S)&= \frac{m-1}{2} \log (n/2) - \log \Gamma(m/2)+ \log C(\S) \\
&+\frac{1}{2} \log \pi +O(1/\sqrt n)
\label{eq-r1}
\end{align}
and the corresponding asymptotic average minimax redundancy is 
\begin{align}
\nonumber
\R_n^{\rm asymp}(\S)&= \frac{m-1}{2} \log (n/2 e) - \log \Gamma(m/2)+ \log C(\S) \\
& +\frac{1}{2} \log \pi +O(1/\sqrt n)
\label{eq-r1-1}
\end{align}
where, we recall,
$$
C(\S)=\Dir(\S; {\bf 1/2})=
\frac{1}{B({\bf 1/2})} \int_{\S} \frac{d\vecth}{\sqrt{\theta_1 \cdots \theta_m}}
$$
as defined above in (\ref{eq-cab}) with $C(\Theta)=1$.
\end{theorem}

We observe that $R^*_n(\S)$ and $\R_n^{\rm asymp}(\S)$ differ approximately
by $\frac{m-1}2$.  This fact 
should be compared with a 
general results of \cite[Theorem 6]{ds04} where it was proved that for 
a large class of sources
$$
|\R_n(\S)-\R_n^*(\S)| \le c_n(\S)
$$
where
$$
c_n(\S)=\sup_{P\in \S}\sum_{x^n} P(x_1^n) \lg \frac
{\sup\limits_{P\in\S} P(x_1^n)}{P(x_1^n)}.
$$
Actually, for binary memoryless sources $c_n(\S) \le 1$ and 
$c_n(\S)\le m-1$ for $m$-ary memoryless sources (\cite[Lemma 8]{ds04} extends 
directly to the $m$-ary case).

In Theorem~\ref{th-r*} we assumed that $m$ is fixed to avoid complications
with constrains $\S_m$ that may depend on $m$. 
In this paper we present our results for large $m=o(n)$. 
While the leading terms, especially for the maximal minimax redundancy,
were known before (see \cite{alon02,alon04,shamir06,shamir06b,sw12,xb97,xb00}), 
our results are derived in a novel way that allows us to apply the methodology
to obtain in Theorem~\ref{cor1} best redundancy results for monotone 
distributions over large alphabet (see \cite{shamir13}).



\begin{theorem}
\label{th-rbar}
Consider a memoryless unconstrained source $\Theta$ with
$m =o(n)$. Then 
the unconstrained maximal redundancy is 
\begin{align}
\label{eq-r3-0}
R_n^*(\Theta)&= \frac{m-1}{2} \log\left(\frac{e\, n}{m}\right) + \frac 12(1-\log 2) 
+O(1/m) +O(m^{3/2}/\sqrt{n}). 
\end{align}
and the unconstrained asymptotic average redundancy becomes
\begin{align}
\label{eq-r3}
\R_n^{\rm asymp}(\Theta)&= \frac{m-1}{2} \log\left (\frac{n}{m}\right) + 
\frac{1}{2}(1-\log 2) 
+O(1/m)+O(m^{3/2}/\sqrt{n}).
\end{align}
\end{theorem}

\begin{remark}
In order to compare Theorems~\ref{th-r*} and \ref{th-rbar} we need to set
$m$ fixed in Theorem~\ref{th-rbar} which means to keep $\Gamma(m/2)$ as
in (\ref{eq-ws1}) and set $C(\Theta)=1$.
\end{remark}

\med
{\bf Monotone Distributions}. As an application of 
Theorems~\ref{th-r*}--\ref{th-rbar}
we provide precise asymptotics for monotone 
distributions which can be viewed as
a special case of the constrained distribution. Indeed, recall
$$
\M=\{ \vecth \in \Theta : \th_1\le \th_2 \le \cdots \le \th_m \}.
$$
Then the unconstrained set $\Theta$ can be divided into $m!$ subsets
$$
\M_\pi=\{  \vecth \in \Theta :
\theta_{\pi(1)} \le  \theta_{\pi(2)} \le ... \le \theta_{\pi(m)} \}
$$
for any permutation $\pi$ of  $\{1, 2,... m\}$.
Furthermore, for all symmetric functionals $f$ it is easy to see that
$$
\int_{\M}  f(\theta_1,..., \theta_m) d\theta_1 ... d\theta_m  =
\frac 1{m!}  \int_\Theta  f(\theta_1,..., \theta_m) d\theta_1 ... d\theta_m.
$$
More precisely, let $D_n(\Theta)$ and $D_n(\M)$  denote the Shtarkov sums 
for the the unconstrained distribution and monotone distribution, respectively.
Then in Section~\ref{sec-cor1} we basically show that (see (\ref{eq-md1}))
$$
D_n(\M)=\frac{1}{m!} D_n(\Theta) +
O\left( (2\pi)^{-\frac{m-1}2} m^{3/2}  n^{\frac{m-2}2} B_m({\bf 1/2})
\exp\left({O\left( \frac {m^{3/2}} {\sqrt n} \right) }\right) \right).
$$
This leads to our next main result regarding the redundancy for monotone
distribution and large alphabet. To the best of our knowledge this is 
the most precise asymptotic expansion (up to $O(1)$ term) for
$m = o(\log n/\log\log n)$ (cf. \cite{shamir13}).

\begin{theorem}
\label{cor1}
Consider a class of monotone distributions $\M$. Then 
$R_n^*(\M)=R_n^*(\Theta)-\log m! +O(m!\, m^{3/2}/\sqrt{n})$.
In particular, for $m=o(n)$ we have
\begin{align}
R^*_n(\M)&=
 \frac{m}{2} \log\left(\frac{n}{m^3}\right) + \frac{3}{2}m \log e +
\frac{1}{2} \log \frac{2\pi m}{e} + \frac 12(1-\log 2) \nonumber \\
&+O(1/m) +O(m!\, m^{3/2}/\sqrt{n}) 
\label{eq-mono}
\end{align}
for large $n$.
\end{theorem}
 

\med
{\bf Sequential Probability Assignment}.
Now, we are ready to present our probability assignment algorithm.
We start with formula (\ref{eq-m2}) on the mixture $M_n(x^n)$. 
Then we observe that $M_n(x_{n+1}|x^n)=M_n(x^{n+1})/M_n(x^n)$.
For example, if assume that $x_{n+1}$ symbol is $i\in \A$. Thus
$$
M_n(x^{n+1})= \frac{ B(k_1+1/2, \cdots, k_i+3/2, \cdots , 
k_m+1/2)}{C(\S) \cdot B({\bf 1/2})} 
$$
$$
\cdot  \Dir(\S; k_1+1/2, \cdots , k_i+3/2, \cdots , k_m+1/2). 
\label{eq-m3}
$$
Using the functional equation
of the gamma function, namely $\Gamma(x+1)=x \Gamma(x)$
allows us to write a simple sequential update algorithm that we present next.

\begin{corollary}
\label{th-algo}
Suppose that $m$ is fixed and that $\S\subseteq \Theta$ is a convex polytope.
Let $N_i(x^n)$ be the number of symbol $i$ in $x^n$. 
Then
\begin{align}
\label{eq-update1}
M_n(x_{n+1}|x^n)=\frac{N_{x_{n+1}}(x^n)+1/2}{n+m/2} \cdot
\end{align}
$$
\cdot \frac{\Dir(\S; N_i(x^n)+1/2+1(x_{n+1}=i),~ i=1\cdots m)}
{\Dir(\S; N_i(x^n)+1/2,~ i=1\cdots m)}
$$
which is the generalized add-1/2-KT estimator.
\end{corollary}

Observe that for the unconstrained case  
$\Dir(\Theta; N_i(x^n)+1/2+1(x_{n+1}=i),~ i=1\cdots m)=
\Dir(\Theta; N_i(x^n)+1/2,~ i=1\cdots m)=1$, and then our estimation algorithm 
reduces to the KT-estimator, that is,
\begin{align}
\label{eq-update2}
M_n(x_{n+1}|X^n)=\frac{N_{x_{n+1}}(x^n)+1/2}{n+m/2}.
\end{align}

We also observe that for the binary alphabet we recover \cite{stw08} update, namely
$$
M_n(x_{n+1}|x^n)=\frac{N_{x_{n+1}}(x^n)+1/2}{n+1} +
+ (2x_{n+1}-1) \frac{a_1^{N_1(x^n)+1/2}(1-a_1{N_0(x^n)+1/2}}{C(\S) (n+1)}
$$
$$
-(2x_{n+1}-1) \frac{b_1^{N_1(x^n)+1/2}(1-b_1{N_0(x^n)+1/2}}{C(\S) (n+1)}.
$$
where $C(\S)$ is defined in (\ref{eq-cab}).
We should point out that the binary sequences sequential 
probability assignment as above was
derived in \cite{stw08} using a different technique that 
seems to be working only for binary sequences.

\section{Analysis and Proofs}

In this section we prove of our main results Theorems~\ref{th-r*}--\ref{cor1}. 
Some technical details are delayed  till the Appendix.
We start with Theorem~\ref{th-rbar} since shall use some ideas and
calculations of the proof of Theorem~\ref{th-rbar} in the proof of 
Theorem~\ref{th-r*}.

\subsection{Proof of Theorem~\ref{th-rbar}}

We now prove Theorem~\ref{th-rbar} where we assume that $m$ may be large
and $\S=\Theta$. By definition, we have $R_n^*(\Theta) = \log D_n$, where
\[
D_n  = \sum_{\bf k} {n\choose {\bf k}} \prod_{i=1}^m \left( \frac{k_i} n \right)^{k_i}
\]
and the sum is taken over all non-negative integer vectors ${\bf k} = (k_1,\ldots,k_m)$
with $\sum_i k_i = n$. If we set
\begin{equation}
\label{eq-md4}
S_m(n) = {\sum_{\bf k}}' {n\choose {\bf k}} \prod_{i=1}^m
\left( \frac{k_i} n \right)^{k_i}
\end{equation}
in which sum $\sum'$ is taken over all $m$-dimensional
integer vectors ${\bf k} = (k_1,\ldots,k_m)$
with $k_j\ge 1$ ($1\le j \le m$) and $k_1 + \cdots + k_m = n$,
then we can represent $D_n$ as
\[
D_n = (2\pi)^{-\frac{m-1}2}\sqrt n \left( 1 + O\left( \frac 1n \right) \right)
\sum_{r=0}^{m-1} {m\choose r} (2\pi)^{r/2} S_{m-r}(n),
\]
that is, $S_{m-r}(n)$ takes care
of those ${\bf k}$, where precisely $r$ components are zero.

By Stirling's formula we have $ k^k e^{-k} \sqrt{2 \pi k} e^{1/(12k+1)} < k! <
k^k e^{-k} \sqrt{2 \pi k} e^{1/(12k)}$ for all $k\ge 1$. Hence $S_m(n)$
satisfies
\[
{\sum_{\bf k}}' \frac {1}{\sqrt{k_1\cdots k_m }} e^{-\sum_{i=1}^m
\frac 1{12 k_i+1} } < S_m(n) < {\sum_{\bf k}}' \frac {1}{\sqrt{k_1\cdots k_m }}
e^{-\sum_{i=1}^m  \frac 1{12 k_i} }.
 \]
By a standard but tedious analysis (see 
Section~\ref{sec-18-19} of Appendix)  we find the asymptotic relation
\begin{equation}\label{eqmd1}
S_m(n)
= n^{\frac m2 - 1} B_m({\bf 1/2})\left(1 +
O\left( \frac{m^{3/2}}{\sqrt n} \right) \right)
\end{equation}
as well as the upper bound
\begin{equation}\label{eqmd1-2}
S_m(n)
= O \left( n^{\frac m2 - 1} B_m({\bf 1/2}) \right),
\end{equation}
where we recall that $B_m({\bf 1/2}) = B({\bf 1/2}) =
\Gamma\left( \frac 12 \right)^m/ \Gamma\left( \frac m2 \right)$.

Then (\ref{eqmd1}) and (\ref{eqmd1-2}) and the relation $B_{m-r}({\bf 1/2}) =
O\left( (m/\pi)^{r/2} B_{m}({\bf 1/2})  \right)$  imply
\begin{align}
D_n &= (2\pi)^{-\frac{m-1}2} n^{\frac{m-1}2} B_m({\bf 1/2})
\left( 1 + O\left( \frac {m^{3/2}} {\sqrt n} \right) \right)  \nonumber \\
&+ (2\pi)^{-\frac{m-1}2}\sqrt n \sum_{r=1}^{m-1} {m\choose r}
O\left(  (2\pi)^{r/2} B_{m-r}({\bf 1/2}) n^{-r/2}  \right) \nonumber \\
&= (2\pi)^{-\frac{m-1}2} n^{\frac{m-1}2} B_m({\bf 1/2})
\left( 1 + O\left(  \sqrt{\frac mn } \right) \right)^m  \nonumber \\
&+ O\left( (2\pi)^{-\frac{m-1}2} n^{\frac{m-1}2} B_m({\bf 1/2})
\frac {m^{3/2}} {\sqrt n}   \right)  \nonumber \\
&= (2\pi)^{-\frac{m-1}2} n^{\frac{m-1}2} B_m({\bf 1/2})
\exp\left({O\left( \frac {m^{3/2}} {\sqrt n} \right) }\right). \label{eqDnfinal}
\end{align}
Since
\begin{align*}
&\log B_m({\bf 1/2}) = m \log \Gamma(1/2) - \log \Gamma(m/2) 
m)
\end{align*}
we directly obtain the proposed representation (\ref{eq-r3-0})
for $R_n^*(\Theta) = \log D_n$.

\medskip

For the asymptotic average minimax $\R_n^{\rm asymp}(\Theta)$ our starting point is
\begin{equation}
\label{eqmd3}
\R_n^{\rm asymp}(\Theta)=\frac{1}{B(\1/2)} \int_{\Theta} \sum_{\bk} {n \choose \bk}
\vecth^{\bk-\1/2}
\log\left(\frac{\vecth^\bk B(\1/2)}{B(\bk+\1/2)}\right)
\end{equation}
where we write $\vecth^{\bk-\1/2}:=\prod_i \th_i^{k_i-1/2}$.
We need to estimate different parts of the above sum.
We first observe that
\begin{align*}
\sum_{\bk} {n \choose \bk} B(\bk+\1/2) &=
\int_\Theta \sum_{\bk} {n \choose \bk} \vecth^{\bk-\1/2} d \vecth \\
&= \int_\Theta \vecth^{-\1/2}d \vecth = B(\1/2).
\end{align*}
More importantly we notice that
\begin{align*}
\int_\Theta \sum_{\bk} {n \choose \bk} \vecth^{\bk-\1/2} \log \vecth^\bk d \vecth
&= \sum_\bk {n \choose \bk} \sum_{i=1}^m \int_\Theta \vecth^{\bk-\1/2} \log \th_i
d\vecth \\
&=\sum_\bk {n \choose \bk} \sum_{i=1}^m k_i \frac{\partial}{\partial k_i} B(\bk+\1/2)
\int_\Theta \sum_{\bk} {n \choose \bk} \vecth^{\bk-\1/2} \log B(\bk+\1/2) \\
&=\sum_{\bk} {n \choose \bk} B(\bk+\1/2) \log  B(\bk+1/2).
\end{align*}
Thus
\begin{equation}
\nonumber
\R_n^{\rm asymp}(\Theta)=\log B(\1/2)+\frac{1}{B(\1/2)}
\sum_\bk {n \choose \bk} B(\bk+\1/2)
\label{ed-md4}
\cdot
\left(\sum_{i=1}^m k_i  \frac{\partial}{\partial k_i}
B(\bk+\1/2) -\log B(\bk+\1/2)\right).
\end{equation}

To deal with such sums we use the relation between the beta function,
the gamma function, and the psi function  \cite{spa-book}. For example
\begin{align*}
\frac{\partial}{\partial k_i} B(\bk+\1/2) &=
\Psi(k_i+1/2)-\Psi(n+m/2)
\end{align*}
where $\Psi(x)=\Gamma'(x)/\Gamma(x)$.
Asymptotically we have 
\begin{align*}
\Psi(x+1/2)&=\log x +1/(12 x)+O(1/x^3),  \\
\Psi(x+m/2)&=\log x+(m-1)/(2x) +O(1/x^2).
\end{align*}
Using this and Stirling's formula we find
\begin{align*}
\log \Gamma(x + 1/2) &= x\log x - x +\log\sqrt{2\pi} - \frac 1{24x} + O(1/x^2), \\
\log \Gamma(x + m/2) &= x \log x - x + \frac{m-1}2 \log(x + m/2)\\
&+ \log\sqrt{2\pi}+ \left( \frac 1{12} - \frac{m^2}8\right) \frac 1x + O(m^3/x^2)
\end{align*}
leading to 
\begin{align*}
\sum_{i=1}^m k_i \frac{\partial}{\partial k_i} B(\bk+\1/2) -\log B(\bk+\1/2)
& = \frac{m-1}{2} \left(\log (n+m/2)-1-\log (2\pi) \right)  \\
&+ O(m^2/n)+O\left(\sum_i (k_i+1)^{-1}\right).
\end{align*}
Hence we obtain similarly to (\ref{eqmd2})
$$
\sum_{\bf k} {n \choose \bk} \frac{B(\bk+\1/2)}{k_i+1} =
O\left(\frac{\sqrt m B(\1/2)}{\sqrt{n}}\right).
$$

Summing up we arrive at
\begin{align}
\label{eq-ws1}
\R_n^{\rm asmp}(\Theta)&=\frac{m-1}{2} \log (n/2\pi e)
+ \log \frac{\Gamma^m(1/2)}{\Gamma(m/2)} +O(m^{3/2}/\sqrt{n}).
\end{align}
We now use Stirling's formula for $\Gamma(m/2)$ and complete the proof.

\subsection{Proof of Theorem~\ref{th-r*}}\label{secproofTh1}

Recall that $R^*_n(\S)= \log D_n$ where
$$
D_n=\sum_{{\bf k}} {n \choose \bk} \sup_{\th \in \S} \prod_{i=1}^m
\theta_i^{k_i}.
$$
The problem is now that we have to distinguish between the case, where
${\bf k}/n \in \S$ and the case, where ${\bf k}/n \not\in \S$.
If ${\bf k}/n \in \S$ then we have
\[
\sup_{\th \in \S} \prod_{i=1}^m
\theta_i^{k_i} = \prod_{i=1}^m \left( \frac{k_i}n \right)^{k_i}
\]
as in the unconstrained case. If ${\bf k}/n \not\in \S$ then we have
\[
\sup_{\th \in \S} \prod_{i=1}^m
\theta_i^{k_i} = \prod_{i=1}^m \theta_{i,{\rm opt}}^{k_i}
\]
where $(\theta_{i,{\rm opt}})$ is on the boundary of $\S$. 

Let us first assume that $\bk/n \in \S$. Then
\begin{align*}
D_n^{(\S)} &:= \sum_{{\bf k}/n \in \S } {n\choose {\bf k}} \prod_{i=1}^m
\left( \frac{k_i}n \right)^{k_i}  \\
&= \left( \frac n{2\pi} \right)^{\frac 12} C(\S) B({\bf 1/2}) 
\left( 1 + O(1/\sqrt n) \right).
\end{align*}

The sum over ${\bf k}$ for which ${\bf k}/n \not\in \S$ is more difficult to handle. 
But if $\S$ is a
convex polytope  we obtain after some (involved) algebra (see the Appendix)
\begin{equation}
\label{eq-error}
D_n - D_n^{(\S)} = O\left( n^{\frac m2 - 1} \right).
\end{equation}
We just mention here the (trivial) case $m=2$ with 
$\S = \{(\theta,1-\theta) : \theta\in [a,b] \}$.
Then
\[
\sup_{0\le\theta \le 1} \theta^{k_1} (1-\theta)^{n-k_1} = a^{k_1} (1-a)^{n-k_1}
\]
and similar for $bn < k_1 \le n$.
Furthermore, 
\begin{align*}
D_n - D_n^{(\S)} &= \sum_{0\le k_1 < an} {n \choose k_1} a^{k_1} (1-a)^{n-k_1} \\
&+ \sum_{bn <k_1 \le n} {n \choose k_1} b^{k_1} (1-b)^{n-k_1} \\
&= 1 + O(1/\sqrt n) = O(1).
\end{align*}
By using $B(\1/2) = \Gamma(1/2)^m /\Gamma(m/2)$ and $\Gamma(1/2) = \sqrt\pi$
we directly obtain (\ref{eq-r1}).

To prove the second statement of Theorem~\ref{th-r*}, the
starting point for the asymptotic average redundancy is (\ref{eq-d1}), however, we
rewrite it in terms of $\S\subseteq \Theta$
as follows
\begin{equation}
\label{eq-md2}
\R_n^{\rm asympt}(\S)=\frac{1}{B_\S(\1/2)} \int_\S \sum_{\bk} 
{n \choose \bk} \vecth^{\bk-\1/2}
\log\left(\frac{\vecth^\bk B_\S(\1/2)}{B_\S(\bk+\1/2)}\right)
\end{equation}
where we use the short hand notation
\[
B_\S ( \veca ) = \int_{\S} \bx^{\veca -1}\, d{\bx} = \Dir(\S; \veca) B(\veca).
\]
As in the proof of Theorem~\ref{th-rbar} we obtain
\[
\R_n^{\rm asympt}(\S)=\log B_\S(\1/2)+
\sum_\bk {n \choose \bk} \frac{B_\S(\bk+\1/2)}{B_\S(\1/2)}
\left(\sum_{i=1}^m k_i  \frac{\partial}{\partial k_i}
B_\S(\bk+\1/2) -\log B_\S(\bk+\1/2)\right).
\]

Again we split the summation over ${\bf k}$ into several parts.
If ${\bf k}/n \in \S^-$, where $\S^-$ denotes all
points in the interior of $\S$ with distance $\ge n^{-1/2+\varepsilon}$
to the boundary (for some $\varepsilon> 0$), then the saddle point
$\th_i = k_i/n$ of the integrand $\vecth^{\bf k}$ of the integral
of $B_\S( {\bf k} + \1/2 )$ or
$\frac{\partial}{\partial k_i} B(\bk+\1/2)$, respectively,
is contained in $\S^-$. Consequently we find for any $L> 0$
\begin{align*}
B_\S( {\bf k} + \1/2 ) &= B( {\bf k} + \1/2 )\left( 1 + O(n^{-L})\right),\\
\frac{\partial}{\partial k_i}
B_\S(\bk+\1/2) &= \frac{\partial}{\partial k_i} B(\bk+\1/2)\left( 1 + O(n^{-L})\right).
\end{align*}
Hence
\begin{align*}
\sum_{\bk/n \in \S^-} {n \choose \bk} & {B_\S(\bk+\1/2)}
\left(\sum_{i=1}^m k_i  \frac{\partial}{\partial k_i}
B_\S(\bk+\1/2) -\log B_\S(\bk+\1/2)\right) \\
&= \sum_{\bk/n \in \S^-} {n \choose \bk} {B(\bk+\1/2)}
\left(\sum_{i=1}^m k_i  \frac{\partial}{\partial k_i}
B(\bk+\1/2) -\log B(\bk+\1/2)\right) + O(n^{-L}) \\
&= \sum_{\bk/n \in \S^-} {n \choose \bk} {B(\bk+\1/2)}  \cdot
\left( \frac{m-1}2 \log \frac n{2\pi e} + O\left(\sum_{i=1}^m 1/(k_i+1) \right)
\right) + O(n^{-L}) \\
&= \left( \frac{m-1}2 \log \frac n{2\pi e} + O\left(1/\sqrt n \right) \right)
B_\S(\1/2).
\end{align*}

The other parts of the summation over $\bk$ are more difficult to handle.
We explain our approach for $m=2$ and
$\S = \{(\theta,1-\theta) : \theta\in [a,b] \}$. Suppose that
$|k_1 - nb| \le n^{1/2+\varepsilon}$, that is $(k_1/n, 1-k_1/n)$ is at distance
$\le n^{-1/2+\varepsilon}$ from the boundary of $\S$. Here we have
\[
B_\S(k_1+1/2,n-k_1+1/2) = \sqrt{\frac{2\pi}n} \left(\frac {k_1}n\right)^{k_1}
\left(\frac {n-k_1}n\right)^{n-k_1}
 \left(\Phi\left( \frac{nb-k_1}{\sqrt{nb(1-b)}} \right) + O(1/\sqrt n) \right),
\]
where $\Phi(u)$ denotes the normal distribution function. A similar representation
holds for the derivatives  $\frac{\partial}{\partial k_i} B_\S(\bk+\1/2)$.
After some algebra it follows that
\[
\sum_{|k_1 - nb| \le n^{1/2+\varepsilon}} {n \choose \bk} {B_\S(\bk+\1/2)} \cdot
\left(\sum_{i=1}^m k_i  \frac{\partial}{\partial k_i}
B_\S(\bk+\1/2) -\log B_\S(\bk+\1/2)\right) = O(1/\sqrt n).
\]
The summation for $nb + n^{1/2+\varepsilon} < k_1 \le n$ is much easier to handle,
so we skip it.

For dimension $m> 2$ one has to handle multivariate Gaussian approximations.
This is just technical and more involved but there is no substantial problem.
Summing up, in all cases the remainder is of order $O(1/\sqrt n)$.
This completes the proof of Theorem~\ref{th-r*}.

\subsection{Proof of Theorem~\ref{cor1}}
\label{sec-cor1}

The goal is to estimate the sum $D_n = D_n(\M)$:
\begin{align*}
D_n &= \sum_{\bf k} {n \choose {\bf k}} \max_{\th_1\le \th_2\le \cdots \le \th_m} 
\th_1^{k_1} \th_2^{k_2} \cdots \th_m^{k_m} \\
&= \sum_{ {\bf k}\in n\M } {n \choose {\bf k}} {n \choose {\bf k}} 
\prod_{i=1}^m \left( \frac {k_i} n \right)^{k_1} 
+  \sum_{ {\bf k}\not\in n\M } {n \choose {\bf k}} 
\max_{\th_1\le \th_2\le \cdots \le \th_m} 
\th_1^{k_1} \th_2^{k_2} \cdots \th_m^{k_m}  \\
&= D_n^{(\M)} + D_n^{(\Theta\setminus \M)}.
\end{align*}

First let us consider the first part $D_n^{(\M)}$ that 
we (again) partition into two parts.
We set
\begin{align*}
\M^< &= \{ {\bf k} \in n\M : k_1 < k_2 < \cdots k_m  \} 
\end{align*}
and then
\[
D_n^{(\M)} =  D_n^{(\M^<)} + D_n^{(\M \setminus \M^<)}.
\]
By using the fact that the functional 
$\prod_{i=1}^m \left( \frac {k_i} n \right)^{k_1}$ is symmetric it
follows that the unconstrained sum (that we denote by 
$D_n(\Theta)$) is given by
\[
D_n(\Theta) = \sum_{\bf k} {n\choose {\bf k}} 
\prod_{i=1}^m \left( \frac {k_i} n \right)^{k_1} 
= m!\,  D_n^{(\M^<)}  + \sum_{k_i = k_j \ 
\mbox{\small for some $i\ne j$}} 
{n\choose {\bf k}} \prod_{i=1}^m \left( \frac {k_i} n \right)^{k_1}.
\]

Now we have 
\[
\sum_{k_i = k_j \ \mbox{\small for some $i\ne j$}} {n\choose {\bf k}} 
\prod_{i=1}^m \left( \frac {k_i} n \right)^{k_i}
\le {m\choose 2} \sum_{k_1 = k_2} {n\choose {\bf k}} 
\prod_{i=1}^m \left( \frac {k_i} n \right)^{k_i}.
\]
With the help of the approximation for $k_i\ge 1$ and $k_1 = k_2$:
\[
{n\choose {\bf k}} \prod_{i=1}^m \left( \frac {k_i} n \right)^{k_1} 
\approx (2\pi)^{-\frac{m-1}2}\sqrt { \frac{n}{k_1^2 k_3 \cdots k_m} },
\]
we arrive at
\begin{equation}
\label{eq-md5}
{\sum_{{\bf k}}}'  \frac 1{ \sqrt {k_1^2 k_3 \cdots k_m} }  =  
\sum_{k=1}^{n/2} \frac 1k S_{m-2}^{(1)}(n-2k),
\end{equation}
where $S_n^{(1)}(m)$ is defined as follows
\begin{equation}
\label{eq-md4a}
S_m^{(1)}(n) = {\sum_{\bf k}}' \frac {1}{\sqrt{k_1\cdots k_m }},
\end{equation}
with the sum taken over all $m$-dimensional
integer vectors ${\bf k} = (k_1,\ldots,k_m)$
with $k_j\ge 1$ ($1\le j \le m$) and $k_1 + \cdots + k_m = n$.
Using the upper bound (\ref{eqmd1-2}) it follows that
\begin{equation}\label{eqsumk1k2est}
\sum_{k_1 = k_2} {n\choose {\bf k}} \prod_{i=1}^m \left( \frac {k_i} n \right)^{k_1}
= O\left( (2\pi)^{-\frac{m-1}2} \sqrt m\,  n^{\frac {m-2}2} \log n \, B_m({\bf 1/2}) 
\exp\left({O\left( \frac {m^{3/2}} {\sqrt n} \right) }\right)  \right).
\end{equation}

Finally, by using (\ref{eqDnfinal}) we get
$$
D_n^{(\M^<)} = \frac 1{m!} 
(2\pi)^{-\frac{m-1}2} n^{\frac{m-1}2} B_m({\bf 1/2}) 
\exp\left({O\left( \frac {m^{3/2}} {\sqrt n} \right) }\right) 
$$
$$
\times
\left( 1 + O\left( \frac{m^{5/2}\log n}{\sqrt n} \exp\left({O\left( \frac {m^{3/2}} {\sqrt n} \right) }\right)  \right) \right).
$$

For the sum $D_n^{(\M \setminus \M^<)}$ we use a simple estimate
\[
D_n^{(\M \setminus \M^<)} \le (m-1) \sum_{k_1 = k_2} 
{n \choose {\bf k}} \left( \frac {k_i} n \right)^{k_i}
\]
that can upper bounded by (\ref{eqsumk1k2est}).

It remains to consider the sum $D_n^{(\Theta\setminus \M)}$, 
where we sum over integer vectors
$(k_1,\ldots,k_m)$ for which there exists $i$ with $k_i > k_{k+1}$. 
Due to the concavity property
of the term $\prod_i \th_i^{k_i} $ (that has optimum outside of $\M$) 
it follows that the optimum
$(\th_1^{\rm opt}, \ldots, \th_m^{\rm opt}) \in \M$ of 
\[
\prod_{i=1}^m  (\th_1^{\rm opt})^{k_i} = \max_{(\th_1,\ldots, \th_m) \in \M}  
\prod_{i=1}^m  \th_i^{k_i}.
\]
has to be on the boundary of $\M$, so
we either have $\th_1 = 0$ or $\th_j = \th_{j+1}$
for some $j = 1, \ldots, m-1$. Hence we trivially have
\begin{align*}
\max_{(\th_1,\ldots, \th_m) \in \M}  
\prod_{i=1}^m  \th_i^{k_i} &\le
0^{k_1} \max_{\th_2 + \cdots + \th_{m-1} = 1}   \prod_{i=2}^m \th_i^{k_i} \\
& + \sum_{j=1}^{m-1}  \max_{\th_1 + \cdots + \th_{j-1} + 2 \th_j + 
\th_{j+2} + \cdots \th_m = 1}   \prod_{i< j} \th_i^{k_i} \cdot 
\th_{j}^{k_j+k_{j+1}}  \cdot \prod_{i >j+1} \th_i^{k_i}.
\end{align*}
The first part is only non-zero if $k_1 = 0$. 
So it simplifies to the $m-1$-dimensional case.
For the second part we note that
\[
\sum_{k_j + k_{j+1} = K}  {n \choose k_1 \cdots k_m } = 2^K
{n  \choose k_1 \cdots k_{j-1}\, K \,  k_{j+2} \cdots k_{m} }.
\]

Thus, we are led to the optimize
\[
\max_{\th_1 + \cdots + \th_{j-1} + 2 \th_j + \th_{j+2} + \cdots \th_m = 1}   
\prod_{i< j} (\th_i)^{k_i} \cdot 
(2\th_{j})^{K}  \cdot \prod_{i >j+1} \th_i^{k_i} 
= \prod_{i< j} \left( \frac {k_i}n \right)^{k_i} \cdot 
\left( \frac Kn \right)^{K}  \cdot \prod_{i >j+1} \left( \frac {k_i}n \right)^{k_i}.
\]
So this case simplifies to the $m-1$-dimensional case, too.
Summing up we have
\begin{equation}
\label{eq-md1}
D_n^{(\Theta\setminus \M)} 
= O\left( (2\pi)^{-\frac{m-1}2} m^{3/2}  n^{\frac{m-2}2} B_m({\bf 1/2}) 
\exp\left({O\left( \frac {m^{3/2}} {\sqrt n} \right) }\right) \right).
\end{equation}

Putting all together, we find
\begin{align*}
D_n &= \frac 1{m!} (2\pi)^{-\frac{m-1}2} n^{\frac{m-1}2} B_m({\bf 1/2}) 
\exp\left({O\left( \frac {m^{3/2}} {\sqrt n} \right) }\right) \\
&+ O\left( (2\pi)^{-\frac{m-1}2} m^{3/2}  n^{\frac{m-2}2} B_m({\bf 1/2}) 
\exp\left({O\left( \frac {m^{3/2}} {\sqrt n} \right) }\right) \right)
\end{align*}
This completes the proof of Theorem~\ref{cor1}.

\newcommand{\refApp}[1]{Appendix~\ref{#1}}

\section{Appendix}

\subsection{Proof of (\ref{eqmd1}) and \ref{eqmd1-2})}

We suppose that $m\ge 2$ and $n\ge 1$ are integers. 
We first estimate $|S_m(n)- S_m^{(1)}(n)|$ where $S_m(n)$ and $S_m^{(1)}(n)$
are defined in (\ref{eq-md4}) and (\ref{eq-md4a}), respectively.
Since $1-e^{-x} \le x$ we have
\begin{equation}\label{eqSmSmest}
\left|  S_m(n)- S_m^{(1)}(n) \right| \le {\sum_{\bf k}}' 
\frac {1}{\sqrt{k_1\cdots k_m }} 
\sum_{i=1}^m \frac 1{12 k_i} =  \frac m{12} {\sum_{\bf k}}' 
\frac {1}{\sqrt{k_1\cdots k_m }} \frac 1{k_1}
\end{equation}

We note that we have the recurrence
\begin{equation}\label{eqSmnrec}
S_m^{(1)}(n) = \sum_{k=1}^{n-1} \frac 1{\sqrt k}  S_{m-1}^{(1)}(n-k).
\end{equation}
and that 
\[
B({\bf 1/2}) = B_m({\bf 1/2}) = \frac{ \Gamma(1/2)^m }{\Gamma(m/2)} = 
\frac{ \pi^{m/2}}{\Gamma(m/2)}
\]
as well as
\[
B_{m-1}({\bf 1/2}) B(1/2, (m-1)/2) = B_m({\bf 1/2}).
\]
Furthermore, for $m\ge 3$ the function $x \mapsto x^{-\frac 12} 
(n-x)^{\frac{m-3}2}$ is decreasing
that leads to the upper bound
\begin{equation}\label{equpperbound1}
\sum_{k=1}^{n-1} \frac 1{\sqrt k} (n-k)^{\frac{m-3}2} 
\le \int_0^{n} \frac 1{\sqrt x} (n-x)^{\frac{m-3}2} dx = 
n^{\frac{m-2}2} B(1/2, (m-1)/2).
\end{equation}

It is now easy to obtain (\ref{eqmd1-2}) by induction.
Clearly we have
\[
\sum_{k=1}^{n-1} \frac 1{\sqrt{k(n-k)}} \le C_1
\]
with a proper constant $C_1 > 0$.  Thus, (\ref{eqmd1-2}) holds for $m=2$.
Furthermore, by using
(\ref{eqSmnrec}) and (\ref{equpperbound1}) we get inductively (for $m\ge 3$)
\begin{align*}
S_m^{(1)}(n) &\le \sum_{k=1}^{n-1} \frac 1{\sqrt k} 
C_1 (n-k)^{\frac{m-3}2} B_{m-1} ({\bf 1/2}) \\
& \le C_1 n^{\frac{m-2}2} B(1/2, (m-1)/2)B_{m-1} ({\bf 1/2})  \\
& = C_1 n^{\frac{m-2}2} B_m({\bf 1/2}).
\end{align*}
This upper bound also shows  that (\ref{eqmd1}) is true if $m\ge c n^{1/3}$ (for any
positive constant $c > 0$. Thus, it will remain to consider the case $m < c n^{1/3}$.

Another application of the upper bound (\ref{eqmd1-2}) leading to an upper bound of 
\begin{align*}
\sum_{\bf k} \frac {1}{\sqrt{k_1\cdots k_m }} \frac 1{k_1} 
&= \sum_{k=1}^{n-1} \frac 1{k\sqrt k} S_{m-1}^{(1)}(n-k) \\
&= O\left( B_{m-1}({\bf 1/2}) \sum_{k=1}^{n-1}  
\frac 1{k\sqrt k} (n-k)^{\frac{m-3}2}  \right).
\end{align*} 
Since $B_{m-1}({\bf 1/2}) = O\left( \sqrt m \,  B_{m}({\bf 1/2}) \right)$ and 
\[
\sum_{k=1}^{n-1} \frac 1{k\sqrt k} (n-k)^{\frac{m-3}2} = 
O\left( n^{\frac{m-3}2}  \right)
\]
we directly obtain
\begin{equation}\label{eqmd2}
\sum_{\bf k} \frac {1}{\sqrt{k_1\cdots k_m }} \frac 1{k_1} 
= O\left( B({\bf 1/2}) m^{\frac 12}  n^{\frac m2 - \frac 32} \right).
\end{equation}
Together with (\ref{eqSmSmest}) we, thus find
\[
\left|  S_m(n)- S_m^{(1)}(n) \right| = O\left( B({\bf 1/2}) m^{\frac 32}  
n^{\frac m2 - \frac 32} \right).
\]

In order to complete the proof of (\ref{eqmd1}) it remains to show that
\begin{equation}\label{eqmd1-3}
S_m^{(1)}(n)
= n^{\frac m2 - 1} B_m({\bf 1/2})\left(1 + 
O\left( \frac{m^{3/2}}{\sqrt n} \right) \right)
\end{equation}
holds for $m < c n^{1/3}$.

For this purpose we use (again) induction and the simplest form of 
the Euler-MacLaurin formula \cite{spa-book}, namely
\[
\sum_{n=a}^b f(n) = \int_a^b f(x)\, dx + \frac{f(a)+f(b)}2 + 
\int_a^b \left(x- \lfloor x \rfloor - \frac 12 \right) f'(x)\, dx,
\]
where $f(x)$ is continuously differentiable and $a<b$ are integers. 
For example, we obtain the asymptotic relation
\begin{align}
S_2^{(1)}(n) &= \sum_{k=1}^{n-1} \frac 1{\sqrt{k(n-k)}} \nonumber \\
&= \int_1^{n-1} \frac{dx}{\sqrt{x(n-x)}} + O\left( n^{-1/2} \right)
+ O\left( \int_1^{n-1} \frac{dx}{\sqrt{x^3(n-x)}}    \right) 
 \nonumber  \\
&= B(1/2,1/2) + O\left( n^{-1/2} \right)  \label{eqS2n}
\end{align}
that is in accordance with (\ref{eqmd1-3}).

Furthermore we find for $m\ge 3$
\begin{align*}
\sum_{k=1}^{n-1} \frac 1{\sqrt k} (n-k)^{\frac{m-3}2} 
&= \int_1^{n-1} \frac 1{\sqrt x} (n-x)^{\frac{m-3}2} dx  + 
O\left( n^{\frac{m-3}2} \right) \\
&+ O\left( \int_1^{n-1} x^{-3/2}   (n-x)^{\frac{m-3}2} dx   \right) +
 O\left( m \int_1^{n-1} x^{-1/2}   (n-x)^{\frac{m-5}2} dx   \right) \\
&= n^{\frac {m-2}2} B(1/2,(m-1)/2) + O\left( n^{\frac{m-3}2} \right) + 
O\left( \sqrt m\,   n^{\frac{m-4}2} \right) \\
&= n^{\frac {m-2}2} B(1/2,(m-1)/2) \left( 1 + 
O\left( \sqrt{ \frac mn } \right) \right).
\end{align*}

With the help of these expansions we will prove (\ref{eqmd1-3}) by induction on $m$.
For notational convenience we write $O_C(X)$ for a term that is 
absolutely bounded by $\le C\, |X|$, that is,
we specify the implicit constant. 

We already mentioned that (\ref{eqmd1-3}) holds for $m=2$, see (\ref{eqS2n}).
Actually we can prove (\ref{eqmd1-3}) for every fixed $m \ge 2$ 
(we just have to apply the inductive
method a finite number of times). Hence, we can assume that (\ref{eqmd1-3})  holds for 
$m\le m_0$, where $m_0\ge 2$ is a fixed but arbitrary integer.  
Now suppose that $m > m_0$ and that (\ref{eqmd1-3}) holds for $m-1$:
\[
S_{m-1}^{(1)}(n) =  n^{\frac {m-3}2} B_{m-1}({\bf 1/2})\left(1 + 
O_C\left( \frac{(m-1)^{3/2}}{\sqrt n} \right) \right).
\]
By (\ref{eqSmnrec}) we, thus, obtain
\begin{align*}
S_m^{(1)}(n) &= \sum_{k=1}^{n-1} \frac 1{\sqrt k} (n-k)^{\frac {m-3}2} 
B_{m-1}({\bf 1/2})\left(1 + O_C\left( \frac{(m-1)^{3/2}}{\sqrt 
{n-k}} \right) \right) \\
&= n^{\frac {m-2}2} B(1/2,(m-1)/2) B_{m-1}({\bf 1/2}) 
\left( 1 + O\left( \sqrt{ \frac mn } \right) \right) \\
&+ O_C\left( (m-1)^{3/2} n^{\frac {m-3}2} B(1/2,(m-2)/2) B_{m-1}({\bf 1/2}) 
\left( 1 + O\left( \sqrt{ \frac mn } \right) \right)  \right).
\end{align*}
Now note that $(m-1)^{3/2} \le m^{3/2}( 1 - c_1/m)$, where 
the constant $c_1>0$ is certainly $\le \frac 32$.
However, if $m$ is sufficiently large (say $m\ge m_0$) then we can assume that $c_1> 1$.
Furthermore we have
\[
B(1/2,(m-2)/2) B_{m-1}({\bf 1/2}) \le B_m({\bf 1/2})\left( 1 + c_2/m \right),
\]
where $c_2 > \frac 12$. However, if $m$ is sufficiently 
large then we can assume that $c_2 < \frac 23$. 
Consequently we have
\begin{align*}
S_m^{(1)}(n) &= n^{\frac {m-2}2} B_{m}({\bf 1/2})
\left( 1 + O\left( \sqrt{ \frac mn } \right)  +
 O_C\left( \frac{m^{3/2}}{\sqrt n} \left( 1 - \frac {c_1} m\right)\left( 1 + \frac {c_2} m\right) \left( 1 + O\left( \sqrt{ \frac mn } \right) \right)  
\right)  \right).
\end{align*}
Recall that we only have to consider the case, 
where $m < c n^{3/2}$, where $c > 0$ can be arbitrarily chosen.
Since $c_1 > 1$ and $c_2 < \frac 23$ we can choose $c> 0$ properly such that
\[
\left( 1 - \frac {c_1} m\right)\left( 1 + \frac {c_2} m\right) 
\left( 1 + O\left( \sqrt{ \frac mn } \right) \right) \le 1 - \frac 1{4m}
\]
for $m_0 \le m  < c n^{3/2}$. Finally, by adjusting $C$ we also assure that
\[
O_C \left( \frac{m^{3/2}}{\sqrt n}\left( 1- \frac 1{4m} \right) \right) + O\left( \sqrt{ \frac mn } \right) 
= O_C \left( \frac{m^{3/2}}{\sqrt n} \right)
\]
for  $m  < c n^{3/2}$. Hence, (\ref{eqmd1-3}) 
follows for all $m  < c n^{3/2}$ and we are done.

\subsection{Proof of (\ref{eq-error})}
\label{sec-18-19}

As mentioned in Section~\ref{secproofTh1} the analysis of the 
proof of (\ref{eq-error}) is very involved.
We recall that a convex polytope is the intersection of 
finitely many half spaces. Thus, in order to 
cover $\Theta\setminus S$ we just have to consider 
the union of finitely many half spaces.
Furthermore, by concavity of $\prod_i \th_i^{k_i}$ it follows that 
the optimal choice of
$\vecth = (\th_1,\ldots,\th_m)$ has to be on the boundary of $\S$.  
Therefore it is sufficient to consider
just one half space $H$ and take the optimum $\vecth$ on the boundary $\partial H$.

Without loss of generality we can assume that the 
boundary $\partial H$ is  given by the equation
\[
a_1 \th_1 + a_2 \th_2 + \cdots +a_{m-1} \th_{m-1} = 1,
\]
where $a_{m-1} \ne 0$. (If all coefficients are non-zero 
we can use the equation $\th_1 + \cdots +\th_m = 1$
to eliminate one variable.) Of course we only consider 
half spaces with the property that
$H\cap \Theta$ is a $m-2$-dimensional polytope of 
positive $m-2$-dimensional volume.
For simplicity we also assume that $a_j\ne 1$, $1\le j \le a_{m-1}$. 
The other cases can be 
treated similarly but the calculations are more involved.

We then analyze the following sum:
\[
S = \sum_{ {\bf k} \in n (H \cap \Theta) } {n \choose {\bf k} }  
\max_{ \vecth \in \partial H \cap \Theta}  \prod_{i=1}^m   \th_i^{k_i}.
\]
In order to simplify our considerations we replace the sum by an integral, that is,
we consider $k_i$ as continuous variables. Since we are in a fixed dimension and 
we have several monotonicity properties it is easy to verify that this approximation
is justified, even more if we are just interested in upper bounds.

Clearly if ${\bf k} \in n (\partial H \cap \Theta)$ then 
\[
\max_{ \vecth \in \partial H \cap \Theta}  \prod_{i=1}^m   \th_i^{k_i}
= \prod_{i=1}^m \left( \frac {k_i}n \right)^{k_i}.
\]
Let's fix such a situation for some ${\bf k}_0 = (k_{1,0},\ldots, k_{m,0}) 
\in n (\partial H \cap \Theta)$, 
that is, we have
\begin{equation}\label{eqkproperties}
\sum_{i=1}^m k_{i,0} = n \quad \mbox{and}\quad   \sum_{i=1}^{m-1} a_ik_{i,0} = n.
\end{equation}
Furthermore we set $\th_{i,0} = k_{i,0}/n$. 

Suppose now that we are not on the boundary of $\partial H \cap \Theta$.
For this case will we characterize those ${\bf k} \in n (H \cap \Theta)$ for which 
\begin{equation}\label{eqmaxproperty}
\max_{ \vecth \in \partial H \cap \Theta}  \prod_{i=1}^m   
\th_i^{k_i} = \prod_{i=1}^m \th_{i,0}^{k_i}.
\end{equation}
(The boundary case has to be handled in a slightly different 
way but the contribution will be negligible.)
For this purpose we parametrize $\vecth \in \partial H \cap 
\Theta$ by $\th_1,\ldots,\th_{m-2}$ 
so that $\th_{m-1}$ and $\th_m$ are given by
\begin{align*}
\th_{m-1} &= \frac 1{a_{m-1}} - \sum_{i=1}^{m-2} \frac{a_i}{a_{m-1}} \th_i,\\
\th_{m} &= \frac {a_{m-1}-1}{a_{m-1}} + \sum_{i=1}^{m-2} 
\frac{a_i-a_{m-1}}{a_{m-1}} \th_i.
\end{align*}
Since our optimum is assumed to be not on the boundary of  
$\partial H \cap \Theta$ and by the concavity property
of the mapping $\vecth \mapsto \sum_i k_i \log \th_i$
it follows that ${\bf k}$ has to satisfy (besides 
(\ref{eqkproperties})) the system of equations
\[
\frac {\partial}{\partial \th_i} \sum_{i=1}^m k_i \log \th_{i,0} = 
\frac{k_i}{\th_{i,0}} - \frac{k_{m-1}}{\th_{m-1,0}} \frac{a_i}{a_{m-1}} + 
\frac{k_{m}}{\th_{m,0}} \frac{a_i-a_{m-1}}{a_{m-1}} = 0, \quad 1\le i \le m-2.
\]
Clearly $k_i = k_{i,0}$ satisfy this system of equations. 

We set $\ell_i = k_i- k_{i,0}$, $1\le i \le m$. 
Then we certainly have $\sum_i \ell_i = 0$
and 
\[
\frac{\ell_i}{\th_{i,0}} - \frac{\ell_{m-1}}{\th_{m-1,0}} \frac{a_i}{a_{m-1}} + 
\frac{\ell_{m}}{\th_{m,0}} \frac{a_i-a_{m-1}}{a_{m-1}} = 0, \quad 1\le i \le m-2,
\]
that is, we have $m-1$ homogeneous equations for $m$ variables.
It is easy to check that the one dimensional solution can be parametrized by $\ell_m$:
\begin{equation}\label{eqli}
\ell_i = \ell_m \frac{\th_{i,0}}{\th_{m,0}} (1-a_i), \quad 1 \le i \le m-1.
\end{equation}
By the way, if we set $a_m = 0$ then this is also true for $i = m$.
Summing up, we have (\ref{eqmaxproperty}) if and only if 
$k_i = k_{i,0} + \ell_i$, $1\le i \le m$,
where $\ell_i$ are given by (\ref{eqli}).

Since we have
\begin{align*}
{n \choose {\bf k}} \prod_{i=1}^m \th_{i,0}^{k_i} &= 
{n\choose {\bf k}_0} \prod_{i=1}^m \th_{i,0}^{k_{i,0}} 
\prod_{i=1}^m \prod_{j=1}^{\ell_i} \frac{k_{i,0}}{k_{i,0} + j } \\
&\sim (2\pi)^{-\frac{m-1}2} \sqrt{ \frac n{k_{1,0} \cdots k_{m,0} } } 
\exp\left( - \sum_{i=1}^m  \frac{\ell_1^2} {2 k_{i,0} }  \right)
\end{align*}
for $k_i = k_{i,0} + \ell_i$, $1\le i \le m$, where $\sum_i \ell_i = 0$, 
and 
\begin{align*}
\sum_{i=1}^m  \frac{\ell_1^2} {2 k_{i,0} } &= \frac{\ell_m^2 n }{2 k_{m,0}^2} 
\sum_{i=1}^m (1-a_i)^2 \th_{i,0} \\
&\ge \frac{\ell_m^2 n }{2 k_{m,0}^2}  \min_i (1-a_i)^2
\end{align*}
we can integrate over $\ell_m$ and obtain in a first step an upper bound for 
\[
S = O\left( (2\pi)^{-\frac{m-1}2}  \int_{K}   
\sqrt{ \frac {k_{m,0}}{k_{1,0} \cdots k_{m-1,0} } } \,  d{\bf k}_0 \right).
\]
Here $K$ denotes the set of all $(k_{1,0},\ldots, k_{m-2,0})$ 
such that ${\bf k}_0 = (k_{1,0},\ldots, k_{m,0}) \in n( \partial H \cap \Theta)$, 
where
\begin{align*}
k_{m-1,0} &= \frac n{a_{m-1}} - \sum_{i=1}^{m-2} \frac{a_i}{a_{m-1}} k_{i,0},\\
k_{m,0} &= \frac {a_{m-1}-1}{a_{m-1}} n + 
\sum_{i=1}^{m-2} \frac{a_i-a_{m-1}}{a_{m-1}} k_{i,0}.
\end{align*}
This directly leads to the upper bound
\[
S = O\left(  n^{\frac{m-2}2} \right)
\]
as proposed in (\ref{eq-error}).

In order to make the arguments completely rigorous we have to 
consider also hyperplanes, where
some of the coefficients $a_i$ equal $1$ and to argue that we 
can estimate sums by integrals.
However, this is just a technical difficulty and not a substantial problem.


\begin{thebibliography}{999}



\bibitem{bernardo}
J. Bernardo,
Reference Posterior Distributions for Bayesian Inference,
{\it J. Roy. Stat. Soc. B.}, 41, 113--147, 1979.

\bibitem{clarkebarron}
B. S. Clarke and A. R. Barron,
Asymptotic of Bayes Information's Theoretic Asymptotics of Bayes Methods,
{\it IEEE Trans. Inform. Theory},  36,  453-471, 1990.

\bibitem{ct}
T. Cover and J.A. Thomas,
{\it Elements of Information Theory},
John Wiley \& Sons, New York 1991.

\bibitem{davisson73}
L. Davisson,
Universal Noiseless Coding,
{\it IEEE Trans. Inform. Theory}, 19, 783--795, 1973.

\bibitem{ds04}
M.~Drmota and W.~Szpankowski,
Precise Minimax Redundancy and Regrets,
{\it IEEE Trans. Information Theory}, 50, 2686--2707, 2004.



\bibitem{flajolet99}
P. Flajolet,
Singularity analysis and asymptotics of Bernoulli sums,
{\it Theoretical Computer Science}, 215, 371--381, 1999.

\bibitem{gallager}
R. Gallager,
{\it Information Theory and Reliable Communication},
John Wiley \& Sons, New York, 1968.



\bibitem{js99}
P. Jacquet and W. Szpankowski,
Entropy Computations via Analytic Depoissonization,
{\it IEEE Information Theory}, 45, 1072-1081, 1999.
Entropy Computations via Analytic Depoissonization,
{\it IEEE Information Theory}, 45, 1072-1081, 1999.

\bibitem{kt83}
R. Krichevsky and V. Trofimov,
The Performance of Universal Coding,
{\it IEEE Trans. Information Theory}, 27, 199-207, 1983.

\bibitem{alon04}
A. Orlitsky and N. P. Santhanam,
Speaking of Infinity,
{\it IEEE Transactions on Information Theory}, 50, 2215-2230, 2004

\bibitem{alon02}
A. Orlitsky, P. Santhanam, and J. Zhang
Universal Compression of Memoryless Sources over Unknown Alphabets,
{\it IEEE Trans. Information Theory}, 50, 1469-1481, 2004.

\bibitem{rissanen96}
J. Rissanen,
Fisher Information and Stochastic Complexity,
{\it IEEE Trans. Information Theory}, 42, 40--47, 1996.


\bibitem{shtarkov87}
Y. Shtarkov,
Universal Sequential Coding of Single Messages,
{\it Problems of Information Transmission}, 23, 175--186, 1987.
{\it Problems of Information Transmission}, 23, 175--186, 1987.



\bibitem{shamir06}
G. Shamir,
Universal Lossless Compression with Unknown Alphabets - The Average Case
{\it IEEE Transactions on Information}, 52, 4915-4944, 2006.

\bibitem{shamir06b}
G. Shamir,
On the MDL Principle for i.i.d. Sources With Large Alphabets,
{\it IEEE Transactions on Information}, 52, 1939-1955, 2006.

\bibitem{shamir13}
G. Shamir,
Universal Source Coding for Monotonic and Fast Decaying Monotonic Distributions,
{\it IEEE Transactions on Information Theory}, 59, 7194-7211, 2013.

\bibitem{stw08}
G. Shamir, T. Tjalkens and F. Willems,
Low-Complexity Sequential Probability Estimation
and Universal Compression for Binary Sequences
with Constrained Distributions,
{\it ISIT}, 2008.
with Constrained Distributions,
{\it ISIT}, 2008.

\bibitem{spa98}
W. Szpankowski,
On Asymptotics of Certain Recurrences Arising in
Universal Coding, {\it Problems of Information Transmission},
34, 55-61, 1998.


\bibitem{spa-book}
W. Szpankowski,
Average Case Analysis of Algorithms on Sequences, Wiley, New York,
2001.

\bibitem{sw12}
W. Szpankowski and M. Weinberger,
Minimax Pointwise Redundancy for Memoryless Models over Large Alphabets,
{\it IEEE Trans. Information Theory}, 58, 4094-4104, 2012.

\bibitem{xb97}
Q. Xie, A. Barron,
Minimax Redundancy for the Class of Memoryless Sources,
{\it IEEE Trans. Information Theory}, 43, 647-657, 1997.

\bibitem{xb00}
Q. Xie, A. Barron,
Asymptotic Minimax Regret for Data Compression, Gambling, and Prediction,
{\it IEEE Trans. Information Theory}, 46, 431-445, 2000.

\end{thebibliography}
\end{document}